\documentclass[12pt]{article}
\newcommand{\non}{\nonumber}
\def\be{\begin{equation}}
\def\ee{\end{equation}}
\def\bea{\begin{eqnarray}}
\def\eea{\end{eqnarray}}
\def\to{\rightarrow}

\def\a{\alpha}
\def\b{\beta}
\def\d{\delta}
\def\e{\epsilon}      
\def\f{\phi}          
\def\g{\gamma}

\def\j{\psi}

\def\m{\mu}
\def\n{\nu}

\def\p{\pi}           
\def\q{\theta}        
\def\r{\rho}          
\def\s{\sigma}        
\def\t{\tau}

\def\L{\Lambda}
\def\O{\Omega}

\def\S{\Sigma}

\def\ca{{\cal A}}

\def\cd{{\cal D}}

\def\cl{{\cal L}}

\def\pa{\partial}                             
\def\ha{\frac12}                              

\begin{document}
\title{{\bf Canonical quantization of Chern-Simons on the light-front}}
\author{L. R. U. Manssur \\
{\normalsize\it Centro Brasileiro de Pesquisas F\'{\i}sicas (CBPF)}\\
{\normalsize\it Departamento de Teoria de Campos e Part\'{\i}culas (DCP)}\\
{\normalsize\it Rua Dr. Xavier Sigaud 150, Urca,}\\
{\normalsize\it 22290-180, Rio de Janeiro, RJ, Brazil}}

\maketitle

\begin{abstract}
{\normalsize
\noindent
By performing the canonical quantization of the Abelian Chern-Simons model on the light-front
(as suggested by Dirac), we clarify some controversies appearing in recent papers that discuss
the relation between the existence of excitations carrying fractional spin and statistics (anyons)
and this model. Properties of the Chern-Simons model on the light-front are investigated in
detail, following the Dirac method for constrained dynamical systems, both for a coupled
complex scalar field as well as for a spinor field.
}
\end{abstract}

\section{Introduction}

\hspace*{\parindent}
Dirac \cite{dirac}, in 1949, noticed that it is possible to quantize dynamical systems
on any space-like surface, in particular a plane wave front moving with the speed of
light, defined by the condition $x^{0}+x^{3}=const$. For simplicity we shall
call that method light-front quantization (LF), in contrast to the usual equal-time
quantization (also referred to as instant form).

LF has some advantages in that seven out of the ten Poincar\'e generators
(in 4 dimensions) are kinematical, while in the instant form only six have this property.
It should also be noticed that non-locality with rescpect to the longitudinal coordinate $x^-$
is expected in the LF, because it happens that the commutators are non-vanishing everywhere
on the light-like lines of the light-front hyperplane \cite{ref10}. In addition, LF always
gives rise to constrained Lagrangians, demanding the use of the well-known Dirac procedure
in order to construct the Hamiltonian.

In 1966, Weinberg \cite{weinberg} obtained Feynman rules in the infinite momentum
frame. In 1970, Kogut and Soper \cite{kogut} proved that those rules correspond to LF
quantization. Even before \cite{fubini}, making $p\to\infty$ was used to derive current
algebra sum rules, and it was noticed that this is the same as using some LF commutators
of the currents.

Recently, interest in LF quantization has been renewed \cite{QCD,ref17} due to the difficulties
in the calculation of non-perturbative effects in the instant form QCD. It occurs that, due
to infrared slavery, the QCD vacuum contains gluonic and fermionic condensates. On the LF,
one obtains a simpler vaccum, often coincident with the perturbative vacuum. This is so because,
for a massive particle on the mass shell, its LF momentum components, $k^{\pm}$, are positive
definite, thus not allowing the excitation of these degrees of freedom on the LF vacuum,
in view of conservation of total longitudinal momentum.

In the context of string theory LF has been used, for example,  in the case of heterotic
string \cite{gross}. Recently, LF has also been used to treat the
chiral Schwinger model \cite{ref15, ref16}.

On the other hand, Chern-Simons models \cite{forte, DJT} have been used to treat planar condensed
matter physics systems, for instance, low temperature superconductivity and quantum Hall effect
\cite{haldane}.

In this paper, we apply LF quantization to Abelian Chern-Simons model in 2+1 dimensions \cite
{prem},
coupled to scalar or spinor fields. As already pointed out, we make use of Dirac procedure
to treat the resulting constrained systems. Our main goal is to clarify the relation between
Chern-Simons systems and the existence, due to commutation of rotations and no {\em a priori}
quantization of angular momentum, of excitations carrying fractional spin and statistics
(anyons) \cite{forte,DJT}. In section 2, we treat Chern-Simons coupled to a complex scalar
field. In section 3, we treat coupling to a fermionic field. In setion 4, we compare both and
discuss our results.

\section{Chern-Simons coupled to scalar fields}

We begin by defining the LF coordinates (in 2+1 dimensions)
\be
x^{\pm}=\frac{x^{0}{\pm} x^{2}}{\sqrt 2}=x_{\mp}.
\ee
We shall adopt $x^{+} \equiv \t$ as our {\em time} coordinate, while $x^{-}$ is our spatial
longitudinal component. The remaining component $x^1$ is the spatial transverse component
also referred to as $x^\perp$. Therefore the greek indices run through $+,-,1$ while
the latin indices take the values $-,1$. The LF coordinates are clearly not related to
the usual ones by any Lorentz transformation.

The theory to be treated here is given by the following Lagrangian density:
\be
\cl=(\cd^\m \f)(\tilde{\cd}_\m \f^*)+ a \e^{\m\n\r} A_\m \pa_\n A_\r,
\ee
where a is a constant, $\f$ is a complex scalar minimally coupled to the Abelian gauge field
$A_\m$ by means of the covariant derivative
\bea
\label{D}
\cd_\m & = & (\pa_\m+iA_\m) \\
\label{Dtil}
\tilde{\cd}_\m & = & (\pa_\m-iA_\m).
\eea
The gauge-invariant Noether current
\be
j^\m=ie(\f^* \cd^\m \f - \f \tilde\cd^\m \f^*)
\ee
is conserved.

The field equations are easily obtained and read as below
\begin{eqnarray}
\label{eqlagrangephi}
\pa_+\pa_-\phi & = & \frac{1}{2}{\cd}_1{\cd}_1\phi-iA_+\pa_-\phi-\frac{i}{2}(\pa_-A_+)\phi \\
\pa_+\pa_-\phi^*& = & \frac{1}{2}{\tilde\cd}_1{\tilde\cd}_1\phi^*+iA_+\pa_-\phi^*+\frac{i}{2}
(\pa_-A_+)\phi^* \\
\label{eqlagrangej+}
2a \pa_-A_1 & = & j^+=i(\f^*\pa_-\f-\f\pa_-\f^*) \\
\label{eqlagrangej-}
2a (\pa_1A_+-\pa_+A_1) & = & j^-=i(\f^*\cd_+\f-\f\tilde\cd_+\f^*) \\
\label{eqlagrangej1}
-2a \pa_-A_+ & = & j^1=-i(\f^*\cd_1\f-\f\tilde\cd_1\f^*),
\end{eqnarray}
after we choose the gauge $A_- \approx 0$.

The $\f$-field satisfy null boundary conditions at spatial infinity $(x^-,x^1)$, while the gauge
field satisfy anti-periodic boundary conditions at infinite $x^-$ and null at infinite $x^1$.
The anti-periodic boundary conditions are in order to allow non-zero electric charge:
\be
\label{condcontorno}
Q=\int d^2x\;j^+=2a\int dx^1\;[A_1(x^-=\infty,x^1)-A_1(x^-=-\infty,x^1)].
\ee

By calculating the canonically conjugate momenta, $\p$, $\p^*$ and $\p^\m$, we conclude that
this is a constrained system, and the constraints are given by
\begin{eqnarray}
 & &\p^+\approx0\\
\chi^i &\equiv & \p^i-a\e^+ij A_j\approx0 \hspace{1in} i,j=-,1\\
\chi &\equiv & \p-\tilde\cd_-\f^*\approx0\\
\chi^* &\equiv & \p^*-\cd_-\f\approx0.
\end{eqnarray}
The canonical Hamiltonian is obtained in the usual way (after integrating by parts):
\be
\label{Hc}
H_c=\int d^2 x \;\left[(\cd_1 \f)(\tilde\cd_1 \f^*)-A_+ \O \right],
\ee
where
\be
\label{omega}
\O=i(\p\f-\p^*\f^*)+a\e^{+ij} \pa_i A_j + \pa_i \p^i.
\ee

Now, we apply the Dirac procedure. Demanding the persistence in time of the
constraints,
we obtain $\O \approx 0$. So, we end up with two first class constraints,
$\p^+\approx0$ and $\O\approx0$, wich generate the gauge transformations. The others are
second class. The pair $A_+$, $\p^+$ decouples, so we only have to deal with $\O$.
We consistently add the constraint $A_- \approx 0$ (thus justifying our choice on the
field equations).

We are ready to define the Dirac brackets,
\be
\{f,g\}_D = \{f,g\}-\int d^2 u\;d^2
v\;\{f,T_m(u)\}\;C^{-1}_{mn}(u,v)\;\{T_n(v),g\},
\ee
where $T_m$ stands for one of the constraints $\chi^i$, $\chi$, $\chi^*$, $A_-$ and $\O$, and
$C_{mn}(x,y)= \{T_m(x),T_n(y)\} $.
The constraint matrix is inverted and we obtain for $C^{-1}_{mn}(x,y)$
\be
\left(
\begin{array}{cccccc}
0 & -4 a{{\partial}^{x}}_{-} & 0   & 0   & 0   & 0  \\
4 a {{\partial}^{x}}_{-} & [\phi^{*}(x)\phi(y)+\phi(x)\phi^{*}(y)] & {2ai}\phi(x)
& - {2ai}\phi(x)^{*} & 0 & -{4 a} \\
0 & {2ai}\phi(y) & 0 & (2a)^2 & 0 & 0 \\
0 & - {2ai}\phi^{*}(y) & (2a)^2 & 0 & 0 & 0 \\
0 & 0 & 0 & 0 & 0 & 2(2a)^2 \\
0 & -{4 a} & 0 & 0 & 2(2a)^2 & 0 \\
\end{array}
\right)
\frac{K(x-y)}{({2a})^2},
\ee
where
\be
K(x-y) = -\frac{1}{4}\e(x^--y^-)\d(x^1-y^1),
\ee
being $\e(x)$ the signal function. This function is antiperiodic at infinite $x^-$ as long
as the above discussed boundary conditions are concerned.
The constraint $\O \approx 0$ gives, in accordance with the field equations,
\be
A_1=\frac{1}{4a}\int d^2y \e(x^--y^-)\d(x^1-y^1) j^+(y),
\ee
and the same boundary condition considerations apply. With $A_-$ and $A_+$ eliminated in
view of the constraints, this leaves only $\f$ and $\f^*$ as independent fields.

By eliminating the constraints, we obtain self-consistently the LF Hamiltonian
\be
\label{Hlf}
H^{l.f.}=\int d^2x(\cd_1\f)(\tilde\cd_1\f^*).
\ee
This Hamiltonian has non-local features as discussed in the Introduction.
We also obtain easily:
\begin{eqnarray}
\label{fifilf}
\{\f,\f\}_D & = & \{\f^*,\f^*\}_D=0 \\
\label{fifi*lf}
\{\f,\f^*\}_D & = & \{\f^*,\f\}_D=K(x-y) \\
\label{fipilf}
\{\p,\f\}_D & = & \{\p^*,\f^*\}_D=-\frac{1}{2}\d^2(x-y)\\
\{\p,A_1\}_D & = & -\frac{i}{4a}[-4\p(x)K(x-y)+\f^*\d^2(x-y)] \\
\label{pdfia1}
\{\f,A_1\}_D & = & \frac{i}{2a}[\f(y)-2\f(x)]K(x-y) \\
\{A_1,A_1\}_D & = & \frac{1}{(2a)^2}[\f(x)\f^*(y)+\f^*(x)\f(y)]K(x-y).
\end{eqnarray}
When checking the self-consistency, we try to recover the field equations using this
brackets. So, we are led to define
\be
\ca_+ \equiv -\frac{1}{4a}\int d^2y \; \e(x^--y^-)\d(x^1-y^1) j^1(y),
\ee
that is identical to $A_+$. This means that, although decoupled throughout the Dirac procedure,
this component must obey the above definition. At the end, we recover the field equations,
having done $A_- = 0$ as a gauge condition, either in Lagrangian or Hamiltonian
formalism.

We are ready to outline the construction of the Fock states. The Dirac bracket are promoted
to commutators,
\bea
\label{lfcomut}
[\f,\f^*] & = & iK(x-y) \\
\label{lfcomutfifi}
[ \f , \f ] & = & 0.
\eea
In order to satisfy them, we use the momentum space expansions
\begin{eqnarray}
\label{expansao}
\f & = & \frac{1}{2\p}\int d^2k \frac{\q(k^+)}{\sqrt{2k^+}} \left[a(k^+,k^1;\t)
e^{-i(k^+x^--k^1x^1)}
+b^{\dag}(k^+,k^1;\t) e^{i(k^+x^--k^1x^1)}\right] \\
\f^* & = & \frac{1}{2\p}\int d^2k \frac{\q(k^+)}{\sqrt{2k^+}} \left[ a^{\dag}(k^+,k^1;\t)
e^{i(k^+x^--k^1x^1)} + b(k^+,k^1;\t) e^{-i(k^+x^--k^1x^1)}\right],
\end{eqnarray}
where $d^2k=dk^+dk^1$ and $\q$ is the Heaviside step function, and with
\be
[a(k), a^{\dag}(k^\prime)]_{\t=\t'}=[b(k), b ^{\dag}(k^\prime)]_{\t=\t'}= \d^2(k-k^\prime)=
\d(k^+-k^{+\prime})\d(k^1-k^{1\prime}),
\ee

If we notice that $A_1$ is a pure gauge,
\bea
A_1 & = & \pa_1\L \\
\L(x) & = & \frac{1}{8a}\int d^2y \; \e(x^--y^-)\e(x^1-y^1)j^+(y),
\eea
then we may define
\be
\hat\f = e^{i\L}\f,
\ee
and rewrite the Hamiltonian as a free one in the field $\hat \phi$
\be
H=\int d^2x (\pa_1\hat\f)(\pa_1\hat\f^*).
\ee
The $\hat \phi$-field does not have a vanishing Dirac bracket with itself,
\begin{eqnarray}
\label{fihatfihat}
[\hat\f(x),\hat\f(y)] & = & e^{\a(x,y)} [e^{i\L(x)},e^{i\L(y)}]\f(y)\f(x)+ \non \\
 & & \hspace{.5in} + \frac{1}{2}\left[e^{i\L(x)}e^{i\L(y)}\f(y^-,x^1)\f(y)
(1-e^{\a(x,y)})-(x\leftrightarrow y)\right]
\end{eqnarray}
where
\be
\a(x,y)=-\frac{i}{8a}\e(x^--y^-)\e(x^1-y^1).
\ee
This is the equivalent on the LF of the equal-time graded commutation relations,
exhibiting manifestly fractional statistics. Due to the boundary conditions, though, the
phase factor $\L$ is single-valued on the LF, while its equal-time equivalent's expression
includes an angle and therefore is multi-valued \cite{DJT}.

It has been shown \cite{prem} that anyonicity seems not to be related to rotational anomaly
(the latter being a gauge artifact), but rather to the dynamics of the CS system, as
given by the Hamiltonian and the Dirac brackets.

\section{Chern-Simons coupled to spinor fields}

We shall use the following representation of the gamma-matrices in 2+1 dimensions:
\bea
\g^0 & = & \s_1 \\
\g^1 & = & i\s_3 \\
\g^2 & = & i\s_2.
\eea
It will be useful to define the LF components of the gamma-matrices
\be
\g^\pm \equiv \frac{1}{\sqrt{2}} (\g^0 \pm \g^2)
\ee
and the projectors
\be
\L^\pm  \equiv \frac{1}{\sqrt{2}}\g^0\g^\pm.
\ee

Our theory is now defined by (writing the spinorial indices
explicitly for later convenience):
\be
\cl=\frac{i}{2}\left[\j^*_\a (\g^0 \g^\m)_{\a\b} \cd_\m \j_\b - \tilde \cd_\m \j^*_\a
(\g^0 \g^\m)_{\a\b} \j_\b \right] + a \e^{\m\n\r} A_\m \pa_\n A_\r,
\ee
where the definition of the covariant derivative still holds.

The gauge-invariant conserved Noether current now reads
\be
j^\m = \j^*_\a (\g^0 \g^\m)_{\a\b} \j_\b.
\ee
The field equations in the gauge $A_- \approx 0$ are
\bea
\label{eqlagrangevinculo}
\sqrt{2} \; \pa_-\j_1 - i \; \cd_1 \j_2 & = & 0 \\
\label{eqlagrangepsi2}
\sqrt{2} \; \cd_+ \j_2 + i \; \cd_1 \j_1 & = & 0 \\
\label{feqlagrangej+}
2a (\pa_- A_1) & = & j^+ = \sqrt{2} \; \j^*_2\j_2 \\
\label{feqlagrangej-}
2a (\pa_1 A_+ - \pa_+ A_1) & = & j^- = \sqrt{2} \; \j^*_1\j_1 \\
\label{feqlagrangej1}
-2a(\pa_- A_+) & = & j^1 = i (\j_1 \j^*_2 - \j^*_1 \j_2).
\eea
Notice that the first field equation has no time derivatives, turning out to be a constraint
in the Hamiltonian formalism. The fields obey boundary conditions similar to the former case.

We calculate the conjugate momenta, $\p_\a$ and $\p^\m$, obtaining the constraints
\bea
\chi_\a \equiv \p_\a -\frac{i}{2} \j^*_\b(\g^0 \g^+)_{\b\a} & \approx & 0 \\
\p^+ & \approx & 0 \\
\chi^i \equiv \p^i - a \e^{+ij} A_j & \approx & 0.
\eea
(Here $\p_1$ is conjugate to $\j_1$ and $\p^1$ is conjugate to $A_1$).
The canonical Hamiltonian is
\be
H_c= \int d^2 x \left\{  \frac{i}{2} \left[ (\tilde\cd_i \j^*_\a)(\g^0\g^i)_{\a\b} \j_\b -
\j^*_\a (\g^0 \g^i)_{\a\b} (\cd_i \j_\b) \right] - A_+ \O \right\},
\ee
where
\be
\O=i(\p_\a \j_\a - \p^*_\a \j^*_\a) + a \e^{+ij} \pa_i A_j + \pa_i \p^i.
\ee

As before, $\O \approx 0$ is a secondary constraint. Further, the persistence in time of
$\chi_\a^*$ gives another secondary constraint, that we shall call $\S \approx 0$.
$\S$, as we expect, is exactly the first of the field equations, as long as we set
$A_- \approx 0$. This last choice is necessary, again, in order to render the matrix of the
Poisson brackets of the constraints invertible.

So, we define the Dirac brackets, now with $T_m(x)$ being
$\chi_1$, $\chi_2$, $\chi^*_1$, $\chi^*_2$, $\chi^-$, $\chi^1$, $\S$, $\S^*$, $A_-$ and $\O$.
The resulting matrix is then inverted. We have developed a technique to simplify the
inversion of the symmetric $10 \times 10$ matrix. It occurs that the matrix is sparse, so we
can take advantage of this fact, by first determining wich elements of the inverse are
necessarily zero. By doing this, we lower from $55 \times 55$ to $10 \times 10$ the system
to effectively determine, with the latter being already block diagonal.

As before, the pair $\p^+$ and $A_+$ decouples. The constraint $\O = 0$ gives
\be
\label{fA1funcaodej+}
A_1(x) = -\frac{\sqrt{2}}{a} \int d^2y K(x-y) \j_2(y) \j^*_2(y),
\ee
and now, in addition, $\S=0$ gives
\be
\j_1 = -i\sqrt{2}\; \int d^2y K(x-y) \cd^y_1 \j_2(y),
\ee
leaving $\j_2$ and $\j_2^*$ as the only independent degrees of freedom.

The LF Hamiltonian ends up being
\be
H^{l.f.}=\ha \int d^2x \left[ \j^*_2 (\pa_1\j_1) + (\pa_1\j^*_1)\j_2 - 2a A_1 \pa_- A_+ \right],
\ee
and we now calculate the folowing Dirac brackets:
\bea
\{\j_2,\j_2\}_D & = & 0 \\
\label{pdpsi2psi2s}
\{\j_2,\j^*_2\}_D & = & -\frac{i}{\sqrt{2}} \d^2(x-y) \\
\{\j_2, \p_2 \}_D & = & \ha \d^2(x-y) \\
\label{pdpsi2j+}
\{\j_2, j^+ \}_D & = & -i \j_2 \d^2(x-y) \\
\label{pdpsi2A1}
\{\j_2, A_1 \}_D & = & -\frac{i}{a} \j_2(x) K(x-y) \\
\label{pdj+j+}
\{ j^+, j^+ \}_D & = & 0 \\
\label{pdA1A1}
\{ A_1, A_1 \}_D & = & 0.
\eea
Now, the charge density commutes with itself, and eq. (\ref{pdpsi2psi2s}) yields a local
commutator, in contrast with the case of the scalar field. We consistently recover the field
equations in the Hamiltonian formalism.

We can also build up the Fock states here, satisfying
\be
[ \j_2 , \j_2^* ]_+ = \frac{1}{\sqrt 2} \d^2(x-y),
\ee
by using
\be
\j_2 = \frac{1}{2^{^\frac{3}{4}} 2 \p}\int d^2 k \left[ a(k^+,k^1;\t)
e^{-i(k^+x^- - k^1x^1)} + b^{\dag}(k^+,k^1;\t) e^{i(k^+x^- - k^1x^1)} \right],
\ee
with
\be
[a(k), a^{\dag}(k^\prime)]_{\t=\t'}=[b(k), b ^{\dag}(k^\prime)]_{\t=\t'}= \d^2(k-k^\prime)=
\d(k^+-k^{+\prime})\d(k^1-k^{1\prime}).
\ee
The remainder of the construction follows as usual.

We can define a $\hat\j_2$,
\bea
\hat\j_2 & = & e^{i\L}\j_2 \\
\L(x) & = & \frac{1}{8a}\int d^2y \e(x^--y^-)\e(x^1-y^1)j^+(y).
\eea
Now, if we use the brackets above, the equivalent of the graded anti-commutation relation of
the equal time formulation reads in the LF
\be
[\hat\j_2,\hat\j_2]_+=0.
\ee
This is also in contrast with the case of the scalar field.

\section{Discussion and final remarks}

We have treated an Abelian Chern-Simons model minimally coupled to a complex scalar
field and a spinor field separately, using a non-covariant gauge, on the LF.
Using the canonical approach and the Dirac procedure for the resulting constrained
system, we have obtained a non-local Hamiltonian with sixth-order interactions,
and showed the self-consistency of our treatment. The gauge field has been eliminated,
leaving just the 2 matter degrees of freedom. We outlined the construction of the Fock
states, in terms of creation and annihilation operators.

We have defined, in the terms of the equal-time formalism, a dual
description, in wich the the matter fields are multivalued and the
Hamiltonian has been claimed to be a free one. It happens that in the
equal-time formalism this can not be so, because having a multivalued field
inside an integral would force us to choose a branch, in order to
integrate, thus introducing a discontinuity and a delta-like interaction.
On the other hand, on the LF, we do not have multivalued fields, in view of
our boundary conditions. So, our Hamiltonian written in terms of the dual
description field is indeed a free one. This dual description field is
single-valued, but does not have usual statisatics, at least in the case of
the scalar field.

The spinor field has a local Dirac bracket, in spite of the fact that this
was not necessary in view of the LF coordinates, and it also has a vanishing
anti-commutator with itself. What remains to be interpreted is the meaning
of the null anti-commutator on the LF.

The next natural step should be to analyse a model with a self-interaction
potential, leading to vortex-like configurations. The above mentioned
boundary conditions should play a central role in such a case.

\section*{Acknowledgements} The author is deeply indebted to
Prof. Prem P. Srivastava for suggesting the problem, for
exhaustive discussions and careful reading of this paper.
We thank Prof. J. A. Helay\"el-Neto for helpful discussions and
for his encouragement. The author would like to thank Centro
Brasileiro de Pesquisas F\'{\i}sicas (CBPF), and the Brazilian
Agency for Scientific Research (CNPq), for financial support.

\end{document}